\newcommand{\beq}{\begin{equation}}
\newcommand{\eeq}{\end{equation}}
\newcommand{\sh}{{\rm sh}}
\newcommand{\ch}{{\rm ch}}
\newcommand{\id}
 {i\kern.06em\hbox{\raise.25ex\hbox{$/$}\kern-.60em$\partial$}}
\newcommand{\as}{\!\not\!\! A}

\newcommand{\bs}{/\kern-.52em b}
\newcommand{\ds}{/\kern-.52em d}
\newcommand{\qs}{/\kern-.52em s}
\newcommand{\D}{{\cal{D}}}

\newcommand{\p}{\partial}
\newcommand{\yp}{^{\prime}}
\newcommand{\bpsi}{\bar{\psi}}

\newcommand{\dd}
{\kern.06em\hbox{\raise.25ex\hbox{$/$}\kern-.60em$\partial$}}

\newcommand{\ep}{\epsilon}

\documentstyle[12pt]{article}
\date{}
\begin{document}
\title{Induced Parity-Breaking Term at Finite Chemical Potential and Temperature
\footnotetext{\# Corresponding adress}
\thanks{On leave of absence from the Physics Department of Shanghai
University, 201800, Shanghai, China}}
\author{{Feng Sze-Shiang $^{1,2}$, Zhu Dong-Pei $^2$}\\
1. {\small {\it High Energy Section, ICTP, Trieste, 34100, Italy}}\\
e-mail:fengss@ictp.trieste.it\\
1.{\small {\it CCAST(World Lab.), P.O. Box 8730, Beijing 100080}}\\
2.{\small {\it Department of Modern Physics , University of Science
and Technology of China, 230026, Hefei, China}}$^\#$\\e-mail:zhdp@
ustc.edu.cn$^\#$}
\maketitle
\newfont{\Bbb}{msbm10 scaled\magstephalf}
\newfont{\frak}{eufm10 scaled\magstephalf}
\newfont{\sfr}{eufm7 scaled\magstephalf}
\baselineskip 0.2in
\begin{center}
\begin{minipage}{135mm}
\vskip 0.3in
\baselineskip 0.2in
\begin{center}{\bf Abstract}\end{center}
  {We exactly calculated the parity-odd term of the effective action induced by the
fermions in 2+1 dimensions at finite chemical potential and finite temperature.
It shows that gauge invariance is still respected.
A more gerneral class of background configurations is
considered. The knowledge
of the reduced 1+1 determinant is required in order to draw exact conclusions
about the gauge invariance of the parity-odd term in this latter case.
  \\PACS number(s):11.10.Wx,11.30.Er
   \\Key words: parity-odd, chemical potential, temperature}
\end{minipage}
\end{center}
\vskip 1in
Thanks to the exotic mathematical structure and the possible relevence to condensed matter
physics in two space dimensions, Chern-Simons(CS) models have drawn much attention in the past decade.
\cite{s1}\cite{s2}(For a review, see \cite{s3}).The CS term  can be either put in by
hand , or more naturaly, induced
by fermion degrees, as a part of the original (effective) lagrangian.Two properties
of the CS action are fundamental. One is that it is odd under parity transform due
to the presence of three dimensional Levi-Civita symbol. The other is that it is invariant
under {\it small} gauge transforms while non-invariant under {\it large }
gauge transforms (those not to be continuously deformed to unity and thus carrying non-trivial winding
numbers)\cite{s4}. 
In the free spacetime whose topology is trivial, the homotopy group $\pi_3$
is trivial 
in the Abelian case. But there may be nontrivial 
 large gauge transformations  if the gauge
fields are
subject to non-trivial boundary conditions(for a more recent discussion  see\cite{s5}).
In general, if there exists non-trivial $\pi_3$, the quantum theory is consistent only
if the CS parameters are quantized. There then arises a problem:what happens to
the quantized parameters by quantum corrections? In the zero temperature, the induced
CS term is well-understood \cite{s6}-\cite{s10}. 
But at finite temperature,
it was argued \cite{s11}that the coefficient of the CS term in the
effective action for the gauge field should remain unchanged at finite temperature.
Yet,  a naive perturbative
calculation that mimics that at zero temperature leads to a CS term with a parameter
continuously dependent on the temperature\cite{s11}-\cite{s12}. Therefore, the bahavior
under gauge transforms seems to be temperature-dependent. 
The problem of quantum corrections to the CS coefficient induced by fermions at finite
temperature was re-examined in\cite{s13}
, where it was concluded that, on gauge invariance grounds and in perturbation
theory, the effective action for the gauge field can not contain a smoothly renormalized CS coefficient at
non-zero temperarture. Obviously, it is neccessary to obtain some exact result
in order to reconcile the contradiction. More recently, the effective action
of a (0+1) analog of the 2+1 CS system was exactly calculated
\cite{s14}. It shows that in the analog, the exact finite $T$ effective action
, which is non-extensive in temperature,
has a well-defined behavior under a large gauge transformation,{\it
independent
 of the temperature}, even though at any given finite order of 
a perturbation expansion, there is a temperature dependence.
So it implies that the discussions of the gauge invariance
of finite temperature effective actions and induced CS terms
in higher dimensions requires consideration of the full
perturbation series. Conversely, no sensible conclusions
may be drawn by considering only the first finite number
of terms in the expansion. The course of being exactly calculable
is that the gauge field can be made constant
by gauge-transformations. Employing this trick,
Fosco {\it et.al.} calculated exactly the parity breaking part of the
fermion determinant
in 2+1 dimensions with a particular background gauge field.
for both Abelian and non-Abelian caases\cite{s15}\cite{s16},
and the result agrees with that from the $\zeta$-function methed
\cite{s5}. 
More general background gauge fields were also considered\cite{s17}.
All these works show that (restricted to that particular {\it ad hoc}
configuration) gauge invariance of the
effective action
is respected even when large gauge transformations are considered.
\\
\indent The effect of  finite chemical potential should be taken into account
whenever discussing the statistical physics of a grand canonical ensemble.
It was shown that in 1+1 dimensions, the non-zero chemical may contribute
a non-trivial phase factor to the partition function\cite{s18} .
The problem for an arbitrary background in 2+1 dimensions was tackled 
perturbatively in \cite{s19}. As ususal, gauge transform property
of the effective action suffers some temperature-dependence. Therefore,
it is worthwhile considering the problem by exact computation with
some particular background. This is the topic of this paper.\\
\indent As usual, the total effective action $\Gamma(A,m,\mu)$ 
is defined as
\beq
e^{-\Gamma(A,m,\mu)}=\int\D\psi\D\bpsi\exp[-\int^{\beta}_0 d\tau\int d^2x\bpsi
(\dd+ie\as+m-\mu\gamma^3)\psi]
\eeq
We are using Euclidean Dirac matrices in the representation $\gamma_{\mu}=
\sigma_{\mu}$, and $\beta$ is the inverse temperature. It makes no difference
whether the indices are lower or upper. The label 3 refers actually to the
Euclidean time component. The fermion fields are subject to antiperiodic
boundary conditions while the gauge field are periodic. Under parity
transformation, 
\beq
x^1\rightarrow-x^1,x^2\rightarrow x^2, x^3\rightarrow x^3;\psi\rightarrow\gamma^1\psi,
\bpsi\rightarrow -\bpsi\gamma^1; A^1\rightarrow -A^1,A^2\rightarrow A^2,A^3\rightarrow A^3
\eeq
($\gamma$ matrices are kept intact). So only the mass term varies under the
parity transformation. As in \cite{s14}, the parity-odd part  is defined
as
\beq
2\Gamma(A,m,\mu)_{\rm odd}=\Gamma(A,m,\mu)-\Gamma(A,-m,\mu)
\eeq
It is not an easy task to calculate (3) for general configuration of the gauge field.
A particular class of configurations of $A$ for which (3) can be exactly
computed is that
\beq
A_3=A_3(\tau), A_j=A_j(x),j=1,2
\eeq
This class of gauge fields shares the same feature as in the 0+1 dimensions:the time
dependence of the time component can be erased by gauge transformations. Therefore, the Euclidean
action can be decoupled as a sum of an infinite 1+1 actions
\beq
e^{-\Gamma(A,m,\mu)}=\int\D\psi_n(x)\D\bpsi_n(x)\exp\{-\frac{1}{\beta}\sum^{+\infty}_{-\infty}
\int d^2x\bpsi_n(x)[\ds+m+i\gamma^3(\omega_n+e\tilde{A}_3)-\mu\gamma^3]\psi_n(x)\}
\eeq
where $\ds=\gamma_j(\p_j+ieA_j)$ is the 1+1 Dirac operator and 
$\tilde{A}_3$ is the
mean value of $A_3(\tau)$. It is seen that the chemical potential in 2+1 dimensions plays the
role of a chiral potential in 1+1 dimensions. Let us introduce $\Omega_n$ for convenience,
$\Omega_n=\omega_n+e\tilde{A}_3$. Since
\beq
m+i\gamma^3\Omega_n-\mu\gamma^3=\rho_ne^{i\gamma_3\phi_n}
\eeq
where
\beq
e^{2i\phi_n}=\frac{m-\mu+i\Omega_n}{m+\mu-i\Omega_n}
\eeq
and
\beq
\rho_n=\sqrt{(m+\mu-i\Omega_n)(m-\mu+i\Omega_n)}
\eeq
we have therefore
\beq
{\rm det}(\dd+ie\as+m-\mu\gamma^3)=\prod^{+\infty}_{n=-\infty}
{\rm det}[\ds+\rho_ne^{i\gamma_3\phi_n}]
\eeq
Explicitly, the 1+1 determinant for a given mode is a functional integral
over 1+1 fermions
\beq
{\rm det}[\ds+\rho_ne^{i\gamma_3\phi_n}]=\int\D\chi_n\D\bar{\chi}_n
\exp\{-\int d^2x\bar{\chi}_n(x)(\ds+\rho_ne^{i\gamma_3\phi_n})\chi_n(x)\}
\eeq
After implementing a chiral rotation whose Jaccobian is wellknown (the Fujikawa
method applies also to complex chiral parameters), we obtain
\beq
{\rm det}[\ds+m+i\gamma^3(\omega_n+e\tilde{A}_3)-\mu\gamma^3]=J_n{\rm det}
[\ds+\rho_n]
\eeq
where
\beq
J_n=\exp(-i\frac{e\phi_n}{2\pi}\int d^2x \ep^{jk}\p_jA_k)
\eeq
Note that the chiral anomalies, or the Jaccobian$J$, dependes on the
boundary conditions as well. If the sysytem is defined on a torus and the
fields are subject to periodic boundary conditions, for instance
,$A_j(x,y)=A_j(x+L_x,y),A_j(x,y)=A_j(x, y+L_y)$, the trace of $\gamma_5$ in
\cite{s20}
is taken over discrete complete set instead of the continuous
plane waves. Thus the momentum integral $\int
\frac{d^2k}{(2\pi)^2}e^{-k^2}=\frac{1}{4\pi}$ should be replaced by
$\frac{1}{L_xL_y}\sum_{n_1,n_2}\exp[-(\frac{2\pi}{L_x}n_1)^2-(\frac
{2\pi}{L_2}n_2)^2]$.
 Using the formula $\sum^{+\infty}_{n=_\infty}
e^{-\pi
zn^2}=\frac{1}{\sqrt{z}}\sum^{+\infty}_{n=-\infty}e^{-\frac{\pi}{z}
n^2}$\cite{s21}
which holds for any complex $z$ with $Rez>0$. 
we have
\beq
\frac{1}{L_xL_y}\sum_{n_1,n_2}\exp[-(\frac{2\pi}{L_x}n_1)^2-(\frac
{2\pi}{L_2}n_2)^2]=\theta(L_x)\theta(L_y)
\eeq
where
$\theta(L)=\frac{1}{\sqrt{4\pi}}\sum^{+\infty}_{n=-\infty}e^{-\frac{L^2}
{4}n^2}
$. In this case, (12) should be replaced by
\beq
J_n=\exp(-2ie\phi_n \theta(L_x)\theta (L_y)\int_{L_x\times L_y} d^2x
\ep^{jk}\p_jA_k)
\eeq
In the following, we only concentrate on the infinite space case since
 the conclusion on a torus can be obtained by a trivial substitution.
Fortunately, we also have  $\rho_n(m)=\rho_n(-m)$ for finite chemical potential. Thus 
 we have
immediately
\beq
\Gamma_{\rm odd}=-\sum^{+\infty}_{n=-\infty}\ln J_n=
i\frac{e}{2\pi}\sum^{+\infty}_{n=-\infty}\phi_n\int d^2x \ep^{jk}\p_jA_k
\eeq
To calculate $\sum^{+\infty}_{n=-\infty}\phi_n$, we need to compute
$\prod^{+\infty}_{n=-\infty}\frac{m-\mu+i\Omega_n}{m+\mu-i\Omega_n}
$.
  Using the formula $\prod_{n=1,3,5,..}[1-\frac{4a^2}{(2n-1)^2}]=\cos\pi a$
as in \cite{s1}, we have ($a=e\tilde{A}_3$)
\beq
\prod^{+\infty}_{n=-\infty}e^{2i\phi_n}=
\prod^{+\infty}_{n=-\infty}\frac{m-\mu+i\Omega_n}{m+\mu-i\Omega_n}
=\frac{\ch\frac{\beta}{2}(m-\mu)+i\sh\frac{\beta}{2}(m-\mu){\rm tg}\frac{\beta a}{2}}
      {\ch\frac{\beta}{2}(m+\mu)-i\sh\frac{\beta}{2}(m+\mu){\rm tg}\frac{\beta a}{2}}
\eeq
Therefore
\beq
\Gamma_{\rm odd}=\frac{e}{4\pi}\ln
[\frac{\ch\frac{\beta}{2}(m-\mu)+i\sh\frac{\beta}{2}(m-\mu){\rm tg}\frac{\beta a}{2}}
      {\ch\frac{\beta}{2}(m+\mu)-i\sh\frac{\beta}{2}(m+\mu){\rm tg}\frac{\beta a}{2}}
]\int d^2x\ep^{jk}\p_jA_k
\eeq
which is quite different from the perturbative conclusion in \cite{s19}. 
(The formula eq(97) there is for an arbitrary background). \\
\indent Now the low temperature limit can be
obtained. It will
depend on
the
relationship between $m$ and $\mu$.\\ 
(i).If $m>\mu,m+\mu>0$
\beq
\lim_{\beta\rightarrow\infty}\Gamma_{\rm odd}=\frac{e}{4\pi}
\beta(ia-\mu)\int d^2x \ep^{jk}\p_jA_k
\eeq
(ii). If $m-\mu>0, m+\mu<0$,
\beq
\lim_{\beta\rightarrow\infty}\Gamma_{\rm odd}=\frac{e}{4\pi}
\beta m\int d^2x \ep^{jk}\p_jA_k
\eeq
(iii).If $m<\mu,m+\mu>0$
\beq
\lim_{\beta\rightarrow\infty}\Gamma_{\rm odd}=\frac{e}{4\pi}
(-\beta m)\int d^2x \ep^{jk}\p_jA_k
\eeq
(iv).If $m<\mu,m+\mu<0$
\beq
\lim_{\beta\rightarrow\infty}\Gamma_{\rm odd}=\frac{e}{4\pi}
\beta(\mu-ia)\int d^2x \ep^{jk}\p_jA_k
\eeq
(v).If $m=\mu$
\beq
\lim_{\beta\rightarrow\infty}\Gamma_{\rm odd}=\frac{e}{4\pi}
(-\beta m+i\frac{\beta a}{2})\ln\cos\frac{\beta a}{2}\int d^2x
\ep^{jk}\p_jA_k
\eeq
 It vanishes in the high temperature limit. It is obvious that the low temperature
is very sensitive to the values of $m$ and $\mu$, as agrees with the results perturbatively
obtained \cite{s19}\\
\indent Since in the large-$m$ limit (or in the low-density limit), the
parity-odd part dominantes over the effective
action, and the particle number in the ensemble is
$<N>=\frac{1}{\beta}\frac{\p}{\p\mu}\ln Z(\beta,\mu)$, we have from the limits 
(18) and (21)
that the flux should be quantized,
\beq
\Phi=<N> \frac{8\pi\hbar}{e}
\eeq
which implies that each particle carries flux $\frac{8\pi\hbar}{e}$
and thus should be of fractional spin $S_{\otimes}=\frac{1}{8}$.
This is
different from the conclusion in \cite{s22}.\\
\indent The previous results can be extended to a more general class of
configurations
\beq
A_j=A_j(x), A_3=A_{30}(x)+\sum_{n\not=0}A_{3n}e^{i\frac{2n\pi}{\beta}\tau}
\eeq
with $A_j $ are locally defined while $A_{3n}$ is globally defined. $A_{3n}$ are constant. The
$x$-dependence of $A_{30}$ results in the $x$-dependence of
$\rho_n$ and $\phi_n$\cite{s15}. In place of (11), we have in this case
\beq
{\rm det}[\ds+m+i\gamma^3(\omega_n+e\tilde{A}_3)(x)-\mu\gamma^3]=J_n^{\yp}{\rm det}
[\ds^{\yp}+\rho_n(x)]
\eeq
where $\ds^{\yp}=\ds-\frac{1}{2}\id\phi_n(x)\gamma^3 $ which depends on the sign of $m$ and thus
contributes
to $\Gamma_{\rm odd}$. The Jaccobian of the cooresponding chiral transformation is
\beq
J_n^{\yp}=\exp\{-i\frac{e}{2\pi}\int d^2x [\phi_n(x)\ep^{jk}\p_jA_k+
\frac{1}{4}\phi_n(x)\Delta\phi_n(x)]\}
\eeq
Now the contribution to
$\Gamma$ from $J_n^{\yp}$, denoted as 
$\Gamma^J$ is
\beq
\Gamma^J =-\sum^{+\infty}_{n=-\infty}\ln J_n^{\yp}=i\frac{e}{2\pi}\int d^2x[
\sum^{+\infty}_{n=-\infty}\phi_n(x)(\ep^{jk}\p_jA_k+\frac{1}{4}\phi_n(x)\Delta
\phi_n(x))]
\eeq
Now we need to calculate 
\beq
\int d^2x \sum^{+\infty}_{n=-\infty}\phi_n(x)\Delta\phi_n(x)=
-\int d^2x \sum^{+\infty}_{n=-\infty}\p^j\phi_n(x)\p_j\phi_n(x)
\eeq
The equality holds because $a(x)$ is globally defined and periodic.
Denote $z_1=-(m-\mu+ia),z_2=m+\mu-ia$, we have
\beq
\sum_n\p^j\phi_n\p_j\phi_n=m^2\p^ja\p_ja
\sum\frac{1}{(i\omega_n-z_1)^2(i\omega_n-z_2)^2}
\eeq
Suppose that $m\pm\mu\not=0$ so that $z_1, z_2$ do not locate on the imaginary axis.
Now the sum can be evaulated using the method analogous to that employed in
\cite{s23}
\beq
\sum\frac{1}{(i\omega_n-z_1)^2(i\omega_n-z_2)^2}
=\frac{-\beta}{2\pi i}\oint_C\frac{dz}{e^{\beta z}+1}\frac{1}{(z-z_1)^2(z-z_2)^2}
\eeq
The contour $C$ encircles the imaginary axis. Then by deforming $C$ to one consisting
of large arcs $\Gamma_n: \mid z\mid=\frac{2n\pi}{\beta}$ which do not pass 
$i\omega_n$ and the parts circumventing
$z_1$ and $z_2$, we have
$$
\sum\frac{1}{(i\omega_n-z_1)^2(i\omega_n-z_2)^2}$$
\beq
=-\beta[\frac{\beta e^{\beta z_1}}{(e^{\beta
z_1}+1)^2(z_1-z_2)^2}+\frac{2}{(e^{\beta
z_1}+1)(z_1-z_2)^3}
+\frac{\beta e^{\beta z_2}}{(e^{\beta z_2}+1)^2(z_2-z_1)^2}+\frac{2}{(e^{\beta
z_2}+1)(z_2-z_1)^3}]
\eeq
Therefore
$$
\sum \p^j\phi_n\p_j\phi_n
=m^2\p^ja\p_ja
(-\beta)$$
\beq
\times
[\frac{\beta e^{\beta z_1}}{(e^{\beta z_1}+1)^2(z_1-z_2)^2}+\frac{2}{(e^{\beta z_1}+1)(z_1-z_2)^3}
+\frac{\beta e^{\beta z_2}}{(e^{\beta z_2}+1)^2(z_2-z_1)^2}+\frac{2}{(e^{\beta z_2}+1)(z_2-z_1)^3}]
\eeq
Hence we have
$$
\int d^2x \sum^{+\infty}_{n=-\infty}\phi_n(x)\Delta\phi_n(x)
=\beta m^2\int d^2x \p^ja\p_ja$$
\beq
\times	
[\frac{\beta e^{\beta z_1}}{(e^{\beta z_1}+1)^2(z_1-z_2)^2}+\frac{2}{(e^{\beta z_1}+1)(z_1-z_2)^3}
+\frac{\beta e^{\beta z_2}}{(e^{\beta z_2}+1)^2(z_2-z_1)^2}+\frac{2}{(e^{\beta z_1}+1)(z_2-z_1)^3}]
\eeq
From (29) we know that the quardratic part in (27) makes no contribution
to the
parity-odd part, therefore we have
$$
\Gamma^J_{\rm odd}=i\frac{e}{2\pi}\int d^2x
\sum^{+\infty}_{n=-\infty}\phi_n(x)\ep^{jk}\p_jA_k
$$
\beq
=\frac{e}{4\pi}\int d^2x\ep^{jk}\p_jA_k
\ln[\frac{\ch\frac{\beta}{2}(m-\mu)+i\sh\frac{\beta}{2}(m-\mu){\rm tg}\frac{\beta a}{2}}
      {\ch\frac{\beta}{2}(m+\mu)-i\sh\frac{\beta}{2}(m+\mu){\rm tg}\frac{\beta a}{2}}]
\eeq
This is identified as the first approximation of the total parity-odd part in
\cite{s15}in the case $\mu=0$.\\
\indent It is seen from (17) that large gauge invariance is still
respected
at finite chemical potential as at the vanishing chemical potential for the background (4).  
For the more general class (24), (34) is also large gauge invariant, but
in this case,
the second factor in (25) contributes also to the parity-odd part.
Therefore, it is neccessary
to consider this factor in order to investigate whether
the gauge invariance of the total parity-odd part is still preserved 
.\\
\indent In summary, we have in this letter discussed the parity-odd part
of the induced effective action of fermions in 2+1 dimensions. For the class of background
field discussed by Fosco {\it et.al.}, the parity-odd part can also be exactly
calculated and is also gauge invariant. For a more general background
field configuration, the total parity-odd part of the effective action can not be obtained
exactly. The total contribution from the chiral rotation can be exactly
calculated nevertheless. Accordingly, knowledge of the second factor, a {\it massive}
fermion determinant, is required in order to draw rigorous conclusions about the
total effective  action. The generalizations to non-Abelian case as well as to 
higher odd dimensions are  straightforward as in
\cite{s16}\cite{s24}. \\
\indent It is clear that we still have not obtained any exact knowledge for
general background gauge field. Since (17)implies in the low-temperature limit, the
everage value $a$ will multiply the two dimesional flux, it seems reasonable
to make the following conjecture\\
{\bf Conjecture} {\it For a general background gauge field, the
dominant part of the parity-odd
term of the induced effective action in 2+1 dimensions is}
\beq
\Gamma_{\rm odd}
=
\ln[\frac{\ch\frac{\beta}{2}(m-\mu)+
i\sh\frac{\beta}{2}(m-\mu){\rm tg}\frac{1}{2}S_{\rm CS}}
      {\ch\frac{\beta}{2}(m+\mu)-i\sh\frac{\beta}{2}(m
+\mu){\rm tg}\frac{1}{2}S_{\rm CS}}]+\cdots
\eeq
{\it where }
\beq
S_{\rm CS}=\frac{e^2}{4\pi}\int 
d^2x \ep^{\mu\nu\alpha}A_{\mu}\p_{\nu}A_{\alpha}
\eeq
 
\vskip 1in
\underline{\bf Acknowledgement} S.S. Feng is grateful 
to Prof. S. Randjbar-Daemi for
reading the manuscript as well as for 
his invitation to ICTP . This work
was
supported by the Funds for
Young Teachers of Shanghai Education Commitee and in part by the National
Science Foundation of China under Grant No. 19805004 and No. 19775044.\\

\end{document}